\documentclass[11pt,reqno]{amsart}

\usepackage{amsmath,amsfonts}
\usepackage{latexsym,amssymb}
\usepackage{bm}
\usepackage[dvipsnames]{xcolor}
\usepackage{graphicx}

\newcommand{\dd}{\mathtt{d}}
\newcommand{\pt}{\partial}
\newcommand{\grad}{\bm{\nabla}}

\begin{document}

\title{What gauges can be used in applied electromagnetic calculations?}

\author{V. Onoochin}

\begin{abstract}
In the classical electrodynamics, different gauges, {\it i.e.} connections between the electromagnetic potentials, are used. Some of these are quite specific and  intended for calculations in special systems (absence of free charges, etc.). All of these specific gauges are reductions of the Lorenz gauge. However, in addition to this gauge, two more, i.e., the Coulomb and velocity gauges, can be used to describe systems of charges and currents without any restrictions.

It is commonly accepted opinion that these three gauges are equivalent, meaning that the expressions for electromagnetic fields obtained from the potentials defined in these gauges are identical. However, it can be shown that the Coulomb and velocity gauges yield solutions corresponding to `superluminal propagation' of the electric field. Since such a propagation of the electric field has not been observed experimentally and, moreover, is forbidden by special relativity, it can be concluded that calculations in these gauges may yield incorrect results. Therefore, these gauges cannot be used in applied electromagnetic calculations. 
\end{abstract}

\maketitle

\section{Introduction}

In classical electrodynamics, the term `gauge' is used in two meanings, {\it i.e.} as {\it gauge invariance} of the electromagnetic fields \cite{Hist} and as {\it gauge fixing}.

The gauge invariance of the EM fields means that the scalar $\varphi$ and vector ${\bf A}$ potentials, which are used to calculate these fields, can be transformed without changing the values of the electromagnetic fields as
\begin{equation}
	\begin{split}
		\varphi' ({\bf r},t)  =\varphi ({\bf r},t) -\frac{1}{c}\frac{\pt \chi({\bf r},t)}{\pt t}\,,\\
		{\bf A}' ({\bf r},t) = {\bf A}({\bf r},t)  +\bm{\nabla}\chi({\bf r},t) \,.
	\end{split} \label{gt}
\end{equation}
where $\chi({\bf r},t)$ is the arbitrary function of time and coordinates. This principle of gauge invariance is widely used in quantum electrodynamics, but not in classical electrodynamics since Eq.~\eqref{gt} is valid not for any gauge. Indeed, if the first of Eqs.~\eqref{gt} were applied to transform the scalar potential calculated with a fixed Coulomb gauge, the gauge function would be zero or constant in time. The scalar potential in the Coulomb gauge is completely determined by the Poisson equation under given boundary conditions: 
\[
\grad^2\varphi_{\rm C}=-4\pi \rho\,,
\] 
and the only solution for the equation,
\[
\grad^2\left[\varphi_{\rm C}-\frac{1}{c}\frac{\pt \chi}{\pt t}\right]=-4\pi \rho\,,
\] 
is $\pt_t\chi = 0$. In quantum electrodynamics, the boundary conditions which unambiguously determine $\varphi_{\rm C}$ are absent.

In contrary to the gauge invariance, the gauge fixing is one of the steps in the procedure for solving the Maxwell equations and is therefore used much more frequently.

Although the Maxwell equations form the basis of electrodynamics, they are not solved directly, but only by introducing $\varphi$ and ${\bf A}$ (in the Gaussian units),
\begin{equation}
	{\bf E} = -\bm{\nabla}\varphi -\frac{1}{c}\frac{\pt {\bf A}}{\pt t} \,;\quad {\bf H} =
	\left[ \bm{\nabla}\times{\bf A}\right] \,.
\end{equation}
One reason to find solution of the Maxwell equations via the equations for the potentials is historical -- Maxwell presented the solution for the EM waves via the wave equation for ${\bf A}$. 

The second reason is more important: if the wave equations for potentials contain the charge and current densities as sources, the wave equations for the EM fields should contain the derivatives of these quantities. In classical electrodynamics, a rigorous definition of the charge density of a classical electron is difficult to give.  
However, the presence of the derivatives $\rho$ and ${\bf J}$ requires an additional definition of the function representing these quantities. Since the charge distribution inside an electron as an elementary particle is unknown, the derivatives $\rho$ cannot be defined rigorously. Here one may ask: why is the definition $\rho({\bf r},t) = e\delta[{\bf r}-{\bf r}_0(t)]$, where ${\bf r}_0(t)$ describes the law of electron motion, applicable in classical electrodynamics? The validity of this definition is formulated in Ch. 19-1 of~\cite{PP} -- $\rho$ as a function describing the classical electron density may not be defined explicitly, but integrating some electrodynamic quantity with this function must yield a finite result independent of the electron's unobservable parameters. Thus, the delta function for $\rho$ satisfies these requirements -- it is not strictly defined, but its integral has a finite value. 

Therefore, to avoid difficulties associated with the derivatives of $\rho$, it is more correct to solve Maxwell's equations by introducing electromagnetic potentials. Rewriting Maxwell's equations via these quantities, one obtains:
\begin{equation}
\begin{split}
-\nabla^2\varphi - \frac{1}{c}\left(\grad \cdot \frac{\pt {\bf A}}{\pt t}\right) = 4\pi \rho\,;\\
-\nabla^2{\bf A} +  \frac{1}{c^2}\frac{\pt^2{\bf A}}{\pt t^2} - \frac{1}{c}\left(\grad \cdot \frac{\pt \varphi}{\pt t} - 
\grad(\grad\cdot{\bf A})\right) = \frac{4\pi}{c}{\bf J}\,.
\end{split} \label{Meq}
\end{equation}
To solve the above system, it is necessary to separate the unknowns $\varphi$ and ${\bf A}$ in the equations. This is achieved by introducing a connection between the potentials, the so-called 'gauge condition'. Such a connection can be introduced in various ways. First, there are special gauge conditions, such as the scalar potential being zero, $\varphi=0$, or ${\bf r}\cdot{\bf A}$, the Poincar\'{e} gauge, and a more general gauge, $n_\mu A_\mu =0$, the light-cone gauge. However, these gauges have a strong limitation: they are valid either in the absence of free charges in the system (radiation gauge), or retardation effects can be neglected when evaluating electromagnetic fields (Poincar\'{e} gauge).
Accordingly, these gauges have very limited application, since they are introduced to solve specific problems. Thus, without any restrictions on the charges and fields in the system, only three of them can be used, {\it namely} the Coulomb, Lorenz, and velocity gauges. In the vast majority of cases, these gauges are used to find solutions for electromagnetic fields. 

The velocity gauge can be treated as generalization of the first two gauges. It follows from definition of this gauge - it is determined by the condition for the potentials $\varphi^{(v)}$ and ${\bf A}^{(v)}$,
\begin {equation}
\grad \cdot {\bf A}^{(v)} + \frac{c}{v^2}\frac{\pt \varphi^{(v)}}{\pt t} = 0\,,
\label{eq-A1}
\end{equation}
Then the above equations become
\begin{equation}
\left( \grad^2 - \frac{1}{v^2}\frac{\pt^2}{\pt t^2}\right) \varphi^{(v)}= - 4\pi \rho\,,
\label{eq-A2}
\end{equation}
\begin{equation}
\left( \grad^2 - {\frac{1 }{ c^2}}{\pt^2 \over \pt t^2} \right) {\bf A}^{(v)} ({\bf r}, t)
= - {4\pi \over c}{\bf J}({\bf r}, t) - c\left( {1 \over v^2} - {1 \over c^2} \right)  \grad \left({\partial\varphi^{(v)}({\bf r}, t) \over \partial t}\right).
\label{eq-A3}
\end{equation}
It follows from Eq.~\eqref{eq-A2} that the parameter $v$ corresponds to a velocity of the scalar potentials propagation.

The Lorenz gauge is a limiting case of the velocity gauge when $v=c$, and the Coulomb gauge is the limiting case of this gauge when $v=\infty$ (instantaneous interaction). 

In fact, apart from special gauges introduced to solve specific problems and therefore of very limited use, only the Lorenz gauge is used in practical calculations.

Moreover, it is illegal to use the Coulomb and velocity gauges, as they can provide pathological solutions for potentials and, consequently, for fields. Let us demonstrate this in the next sections.

\section{Derivation of `superluminal solution' for the electric field in the Coulomb and velocity gauges.}

The system of Eqs.~\eqref{eq-A2} and \eqref{eq-A3} is solved in the following sequence: first, the solution for $\varphi^{(v)}$ is found that is represented by the retarded integral  
\begin{equation}
\varphi^{(v)}({\bf r},t)=\frac{1}{4\pi}\int \frac{\rho\left({\bf r}',t- |{\bf r}-{\bf r}'|/v\right)}{|{\bf r}-{\bf r}'|}\dd {\bf r}'\,,\label{v-Phi}
\end{equation}
where the retarded time is determined as $t_{ret}=t - |{\bf r}-{\bf r}'|/v$. 

Then $\grad\left(\pt\varphi^{(v)}/\pt t\right)$ is calculated, and this term is inserted into {\it rhs} of \eqref{eq-A3}. After that, a solution for the vector potential is sought.

Significant property of the Coulomb and velocity gauges is that their scalar potentials are assumed to propagate with superluminal velocity. So it is reasonable to verify whether these potentials can give solutions corresponding to superluminal propagation of the electric field. To do it let us  consider the following system: a charge $q$ that had been at rest at p. $P$  until time $t = 0$ and then it started suddenly to move with a constant velocity $u$ along the $x$-axis (Fig. 1). $D$ is the distance between points $P$ and $O$. Due to its simplicity, the potentials of this system, defined in the Coulomb gauge, can be evaluated in the closed form, at least at point $O$, the origin of the coordinates, where a detection of the electric field is being.

\begin{figure}[htbp]
\begin{center}
\includegraphics[bb = 0 0 709 307, scale=0.65]{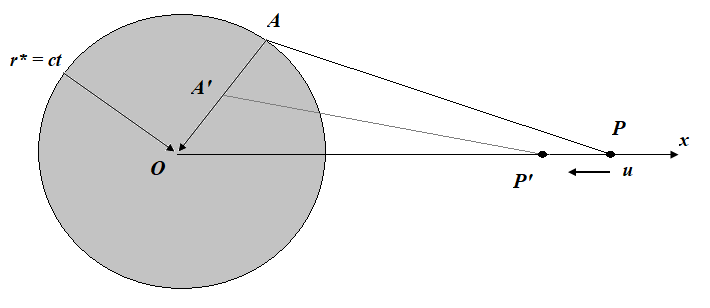}
\caption{A system where calculations of the electric field in the Coulomb gauge predict superluminal propagation.}
\end{center}
\end{figure}
A case of superluminal signal corresponds such a locations of the charge and the field detector that when the EM wave, emitted by the charge at $t=0$, does not reach the observation point $O$ at time $t$, or $D>ct$. Meanwhile, the electric field component ${\bf E}_{\varphi}=\bm{\nabla}\varphi_{\rm C}$ reaches the detector at time $t=0$. But this component is not the only one that reaches the detector at this time. The source of the wave equation for ${\bf A}_{\rm C}$ also propagates with the same (infinite) velocity and at $t=0$ this source must have a non-zero value in the neighborhood of point $O$. This source creates a vector potential and, consequently, an electric field component ${\bf E}_{\bf A}=\pt_t{\bf A}_{\rm C}$. 

In other words, although the vector potential in the Coulomb gauge propagates at a finite speed (of light), the source creating this potential propagates at an infinite speed. As a result, both components of the electric field will be registered by the detector.

However, the detector is expected to show a zero signal at $t=0$ if the two components cancel each other out, and the only component of the electric field in the system is the component created by the local $4\pi{\bf J}/c$ source and propagating at the speed of light. If the ${\bf E}_{\varphi}$ and ${\bf E}_{\bf A}$ components do not cancel each other out, the detector should register a superluminal signal. Let us consider what electric field is created at point $O$ by the potentials of a moving charge, provided that these potentials are calculated in the Coulomb gauge. 

Electric field calculated in this gauge is given as
\[
{\bf E}_{\rm C}(0,t)=\frac{q}{(D-ut)^2} - \frac{1}{c}\frac{\pt {\bf A}_{\rm C}(0,t)}{\pt t}\,.
\]
The vector potential is created by two sources; our aim is to analyze the component of the vector potential ${\bf A}'_{\rm C}$ created by non-local source. Let us rewrite ${\bf E}'_{\bf A}$ as
\begin{equation}
 \frac{1}{c}\frac{\pt {\bf A}'_{\rm C}({\bf R},t)}{\pt t}= \frac{1}{c^2}\frac{\pt }{\pt t}
 \int \frac{1}{|{\bf R}-{\bf r}|}\left[\grad_r\frac{\pt \varphi_{\rm C}}{\pt t} \right]_{ret}\dd {\bf r}=\grad_R\frac{1}{c^2}
  \int \frac{1}{|{\bf R}-{\bf r}|}\left[\frac{\pt^2 \varphi_{\rm C}}{\pt t^2} \right]_{ret}\dd {\bf r}\,,\label{nabl}
\end{equation}
where this retarded time is calculated for $c$, the speed of the vector potential propagation. The indexes $r$ and $R$ at the operator  $\grad$ mean that this operator acts on the variable ${\bf r}$ or ${\bf R}$. All details of the integral transformation are given in Appendix A of~\cite{O}.

The next step is to evaluate integral in the {\it rhs} of~\eqref{nabl} as it is done in~\cite{O}. This integral is transformed from the vector potential, which is a solution to the wave equation. When calculating the  vector potentials ${\bf A}_{\rm C}$ at a given point, it is necessary to collect an account of all the waves emitted by the extended source $\grad (\pt \varphi_{\rm C}/\pt t)$. Let us calculate the value of this integral in a similar manner. 

When the charge was at rest at $t< 0$, value of this integral is equal to zero. When $t>0$, the value of the integral is formed by all waves emitted by the non-local source and approached p. ${O}$ at the time $t$, {\it i.e.} all waves in the sphere of radius $r^\star=ct$ converging to its central point, ${O}$. So when the charge is at p. ${P}$ ($t=0$, $\varphi_{\rm C}$ creates the converging wave on the surface of the sphere), 
\[
\frac{\pt^2\varphi_{\rm C}}{\pt t^2}=\lim_{t\to 0}\frac{\pt^2}{\pt t^2}\frac{q}{\sqrt{[D-x-u t]^2+y^2+z^2}}\,.
\]
where $x,\,y,\,z$ are coordinates of p. ${A}$.

Correspondingly, when the charge is at p. $P'$, $\varphi_{\rm C}$ creates the converging wave on the surface of the sphere with the radius 
$r'=\sqrt{x'^2+y'^2+z'^2}=ct'$. Here, $x',\,y',\,z',\,$ are the coordinate of p. $A'$ on this sphere. The source for this integral is
\[
\frac{\pt^2\varphi_{\rm C}}{\pt t^2}=\frac{\pt^2}{\pt t'^2}\frac{q}{\sqrt{[D-x'-u t']^2+y'^2+z'^2}}\,.
\]
Thus, the integral in the {\it rhs} of Eq.~\eqref{nabl}, written in the spherical coordinates $r\,,\omega,\,\phi$ ($x=r\cos\omega$, and integration over $\phi$ gives $2\pi$), becomes
\begin{equation}
I_{\rm C}=\frac{qv^2}{2 c^2}\int \limits_0^{ct}\frac{r^2\dd r}{r}\int\limits_{-\pi}^{\pi} 
\frac{2\left[D-r\cos\omega-u(t-r/c)\right]^2-r^2\sin^2\omega}{
\left\{\left[D-r\cos\omega-u(t-r/c)\right]^2+r^2\sin^2 \omega\right\}^{5/2}}\sin\omega\dd \omega\,.\label{Int}
\end{equation}
For the case of `superluminal signal' the above integral contains no singularities, and its calculation with respect to the angular variable $\omega$ by means of {\it Mathematica} software yields,
\begin{equation*}
I_{\rm C}=2qu^2c\int \limits_0^{ct} \frac{r\dd r}{[c(D-ut)+ur][c(D-r-ut)+ur]
[c(D+r-ut)+ur]}\,.
\end{equation*}
Evaluation of the above integral does not have difficulties and the result is
\begin{equation*}
I_{\rm C}= \frac{u^2}{c^2}\dfrac{\ln\left[1-\dfrac{c^2t^2}{D^2}\right]}{D -u t }
\end{equation*}
Then the electric field detected at p. $O$ at the instant $t$ is 
\begin{equation}
E_{\rm C, x}=\frac{q}{(D-ut)^2}\left\{ 1-\frac{u^2}{c^2}\ln \left[1-\frac{c^2t^2}{D^2} \right]\right\}\,,
\end{equation}
where it is taken into account that $r_0(t)=vt$ and calculation of gradient of potentials along the $x$ axis is equal to calculation of the partial derivative $\pt_x=\pt_{vt}$.

Now it is necessary to consider calculation of the vector potential in the velocity gauge. This consideration allows to `see' obstacles in attempt to evaluate the integral similar to the integral of Eq.~\eqref{Int} but written in the velocity gauge.

As in the case of the Coulomb gauge, the source of the wave equation for the vector potential ${\bf A}^{(v)}$ is non-local. However, it cannot be distributed over the entire space, but only in the area occupied by the `scalar potential waves' $\varphi^{(v)}$ at time $t$. Let us introduce an auxiliary time variable $\tau$ to describe the motion of the charge ${\bf r}_0=ut$ such that at $\tau=0$ the charge is at the point p. $P$. This allows us to determine the area in which the vector potential waves will converge to p. $O$ over time $[0;\,t]$ (an analog of a sphere of radius $r^\star=ct$ for the Coulomb gauge, Fig.~1). This area is bounded by an axially symmetric surface (the $x$ axis is the axis of symmetry) -- Fig. 2.
\[
\frac{r}{c}+\frac{\sqrt{(D-r\cos\omega)^2+r^2\sin^2\omega}}{v}=t\,.
\]
This area is presented in Fig.~2, labeled by the gray color. 

\begin{figure}[htbp]
\begin{center}
\includegraphics[bb = 0 0 736 508, scale=0.54]{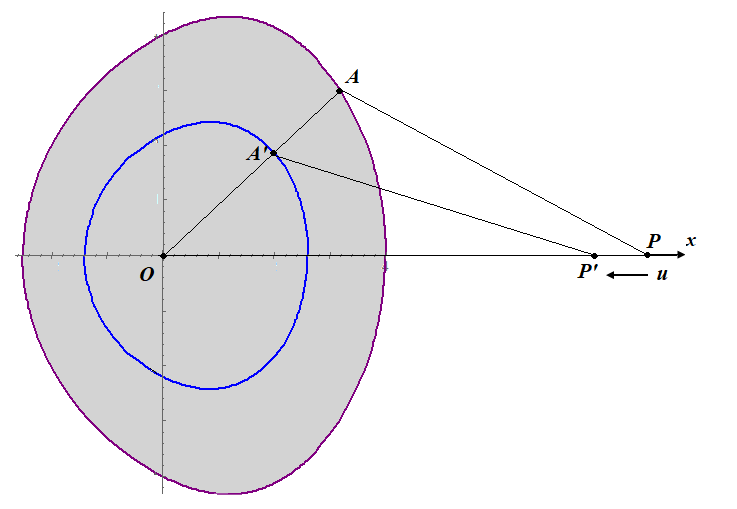}
\caption{A system where calculations of the electric field in the velocity gauge predict superluminal propagation.}
\end{center}
\end{figure}

To sum up the contributions from all points in the named region, where the distributed source $\pt^2 \varphi^{(v)}/\pt t^2$ creates converging waves (to point $O$) by the time $t$, it is convenient to introduce an auxiliary time variable $\tau$, or the time of charge motion towards the detector.
\begin{equation}
\frac{r}{c}+\frac{\sqrt{(D-r\cos\omega-u\tau)^2+r^2\sin^2\omega}}{v}=t-\tau\,. \label{tau}
\end{equation}
Then $t-\tau$ is similar to the retarded time in the Coulomb and Lorenz gauges. 

If $r$ has a complex dependence on the angle $\omega$ and $\tau$, then the term describing the source has a more or less simple form. Since the charge moves along the $x$ axis with a velocity $u=const$, the scalar potential is determined by the expression
\begin{equation}
\varphi^{(v)}	({\bf r}',t)=\frac{q}{\sqrt{(D-x'-ut)^2+[1-(u/v)^2](y'^2+z'^2) } },. \label{phiV}
\end{equation}
Then the complete integral corresponding to $I_{\rm C}$ is
\begin{equation}
I^{(v)}=\frac{q}{4 c^2}\int\limits_{\tfrac{D}{v+u}}^0\dd \tau\int\limits_0^{\pi}
\frac{\pt r^2}{\pt \tau}\frac{\pt^2}{\pt \tau^2}\left[\dfrac{1}{\sqrt{(D-r\cos\omega-u\tau)^2+\left[1-\dfrac{u^2}{v^2}
\right]r^2\sin^2\omega } }\right] \sin\omega\dd \omega\,. \label{IntV}
\end{equation}
where $r$ is the function of $\tau$ and $\omega$ determined by Eq.~\eqref{tau}.

Although this integral cannot be evaluated due to cumbersome dependence of $r$ on $\tau$ and $\omega$, some result can be obtained from the analysis of Eq.~\eqref{IntV}, {\it namely} this integral depends on the factor $u^2$, arising from double differentiation with respect to $\tau$, and can be written as 
\[
I^{(v)}= \frac{q u^2}{c^2} F(u,\,v,\,c\,,t,\,D)\,,
\]
where $F(u,v,c,t,D)$ is some function. This result means that the electric field components, calculated in `superluminal sector' ($D>ct$), do not cancel each the other since the component created by ${\bf A}^{(v)}$ is proportional to $u^2$ but the component created by $\varphi^{(v)}$ does not contain the factor $u$ in the numerator of Eq.~\eqref{phiV}. This means that superluminal propagation of the electric field should also be predicted from calculations of the potentials in the velocity gauge.

\section{On alternative representation of the potentials in the velocity gauge}

Since a complete expression for the vector potential in the velocity gauge is impossible to calculate, some attempts have been made to develop another expression representing ${\bf A}$ in this gauge. These attempts are based on the assumption that the vector potential ${\bf A}^{(v)}$, or at least one of its components, propagates with the same velocity as the scalar potential $\varphi^{(v)}$. Let us consider these attempts in more detail. 

The first work in this direction is by Yang~\cite{Yang-76}, who makes a statement that one component of the vector potential propagates with the speed $v$. In the other words, the vector potential can be decomposed as
\begin{equation}
{\bf A}^{(v)}= {\bf A}_c + {\bf A}_v
\end{equation}
where ${\bf A}_c + {\bf A}_v$ are the components  propagating with the speeds, $c$ and $v$ respectively (instead of $v$ Yang used the notation $\alpha c$ with $\alpha >1$):\newline
{\it From} Eq. (3.15),{\it it is obvious that ${\bf A}_c^{(\alpha)}$ must possess at least one term which must travel at speed $\alpha c$ if $\alpha$ is not one to cancel exactly the gradient of the scalar potential so that the resulting fields always propagate at speed $c$ from the source regions}.

But this statement contains at least two weak point.\newline
1. It is not obvious that one term should propagate with the speed $v$. The propagation speed of the vector potential is uniquely determined by the wave equation $\left[\grad^2-(1/c^2)\pt_t^2\right]$, and this speed is equal to $c$. Although the source for ${\bf A}^{(v)}$ propagates with the speed $v$, this is not the actual propagation speed of the vector potential. This parameter must be determined from the corresponding wave equation. The author~\cite{Yang-76} does not derive a wave equation of the form $\left[\grad^2-(1/v^2)\pt_t^2\right]$ for ${\bf A}^{(v)}$;\newline 
2. As it follows from the arguments presented at the end of Sec.~2, a time derivative of the vector potential do not cancel the gradient of the scalar potential.

Therefore, the author's assumptions cannot be used to derive the final result of~\cite{Yang-76}, the expression for the vector potential in the velocity gauge convenient for further calculations (Eq.~(3.26)). 

A certain progress in deriving the expression for  ${\bf A}^{(v)}$ convenient for calculations is made in~\cite{BC}. The authors rewrite the initial equations for potentials, Eqs.~\eqref{eq-A2} and~\eqref{eq-A3} in the form (Eqs. ~(27) of the cited paper),
\begin{eqnarray}
\left( \grad^2 - \frac{1}{c^2}\frac{\pt^2}{\pt t^2}\right) {\bf A}+\left(\frac{v^2}{c^2}-1\right)\grad (\grad\cdot {\bf A}) = -\frac{4\pi}{c}{\bf J}\,;\label{Av}\\
	\left( \grad^2 - \frac{1}{v^2}\frac{\pt^2}{\pt t^2}\right) \varphi =-4\pi \rho\,. \nonumber 
\end{eqnarray}
In the above system, the potentials are separated.  But the authors do not seek a solution of Eq.~\eqref{Av}. Instead they introduce the dyadic Green function $\Theta_\gamma({\bf x},t||{\bf x'},t')$ which should satisfy the wave equation with the speed of propagation of 'waves of this equation' equal to $v$,
\[
\left( \grad^2-\frac{1}{v^2}\frac{\pt^2}{\pt t^2}\right)\Theta_{\gamma}(x,t||x't')=\delta({\bf x}-{\bf x}')\delta(t-t')
\]
An application of this equation can be found in the transition from Eq.~(30) to Eq.~(31) in~\cite{BC}. But since this Green's function is used to find the vector potential, introducing this dyadic function is equivalent to assuming that one of the components of the vector potential propagates with the speed $v$. Since the Green's function enters Eq.~(30) in the combination $\left(\grad\cdot \Theta_\gamma({\bf x},t||{\bf x'},t')\right)$, this means that the {\it longitudinal component} of the vector potential must propagate with velocity $v$.

Again, the weakness of this assumption -- like Yang's assumption -- is that in reality the wave equation $\left[\grad^2-(1/v^2)\pt_t^2\right]$ is introduced in~\cite{BC} without proper justification. However, the absence of validity of this introduction makes the authors' results incorrect.

Another attempt to solve the problem of calculating fields in the Coulomb and velocity gauges should be analyzed. This attempt is based on the idea of demonstrating that the expressions for the electromagnetic field are the same regardless of the gauge in which the potentials are calculated. If this assumption is correct, then there is no need to calculate the potentials in any gauge other than the Lorenz gauge; it is sufficient to write down the potentials in an electrodynamic system in any gauge and then, if necessary, use the Lorenz--gauge potentials to calculate the electromagnetic fields in that system.

 This idea arises from one property of the potentials, namely, if partial time derivative of any function $\chi$ is added to the scalar potential and the gradient of the same function is added to the vector potential, the expressions for the EM fields do not change (Eqs.~\eqref{gt}). 
 
But these transformations of potentials with an {\it arbitrary} function $\chi$ are valid only if the set of potentials $\varphi,\,\varphi',\,{\bf A},\,{\bf A}'$ is defined in the same gauge. An arbitrary function $\chi$ cannot transform a potential $\varphi_{\rm L}$, defined, for example, in the Lorenz gauge, into a potential $\varphi^{(v)}$ defined in the velocity gauge. This follows from the fact that $\varphi_{\rm L}$ depends on the parameter $c$, the speed of light, and $\varphi^{(v)}$ depends on the parameter $v$, the propagation velocity of the scalar potential in corresponding gauge. Therefore, $\chi$ cannot be arbitrary, but must depend on $v$. Consequently, it is necessary to derive a gauge function. 
 
The most detailed development of this idea is presented in~\cite{Jack}, where the author introduces a function $\chi_C$, which is determined from the scalar potentials in the Lorenz and Coulomb gauges (sec.~III), and the function $\Psi$, which is determined from the vector potentials in the Lorenz and Coulomb gauges (sec.~II). Then the author shows that these functions are equal each the others.  In fact, it is sufficient to conclude that the idea to transform the potentials from one gauge to the other any gauge has solid mathematical background.

However, there is one difficulty in the introduction of $\Psi$. According to Eq.~(2.9) of~\cite{Jack}, the vector potential in the Coulomb gauge can be presented as a sum of the longitudinal and transverse components,
\[
{\bf A}_C={\bf A}_l+{\bf A}_t\,;\quad {\bf A}_l=\bm{\nabla}\Psi\,,\,\, {\bf A}_t=[\bm{\nabla}\times{\bf V}]\,.
\]
But the vector potential in the Coulomb gauge is defined as $\bm{\nabla}\cdot{\bf A}_C=0$. Thus we must have
\[
\left(\grad\cdot{\bf A}_C\right)=\grad \cdot \grad \Psi +\left(\grad\cdot[\bm{\nabla}\times{\bf V}]\right) =\grad^2\Psi = 0\,.
\]
The quantity $\Psi$ is given by Eq.~(4.1) of~\cite{Jack},
\[
\Psi({\bf r},t)=\frac{1}{4\pi c}\int \frac{1}{|{\bf r}-{\bf r}'|}\left[ \int \dfrac{\pt_t \rho[{\bf r}''-{\bf r}_0(t-|{\bf r}-{\bf r}'|/c)] }{|{\bf r}'-{\bf r}''| }\dd {\bf r}''\right]_{ret} 
 \dd{\bf r}'\,,
\]
where ${\bf r}_0(t)$ represents the law of the charge motion. Then  evaluation of action of $\grad^2$ on the above expression gives
\begin{equation}
\begin{split}
\grad^2\Psi ({\bf r},t)= \frac{1}{c} \int \delta({\bf r}-{\bf r}')\left[ \int \dfrac{\pt_t \rho[{\bf r}''-{\bf r}_0(t-|{\bf r}-{\bf r}'|/c)] }{|{\bf r}'-{\bf r}''| }\dd {\bf r}''\right] _{ret}
 \dd{\bf r}'=\\
=\frac{1}{c}\int \dfrac{\pt_t \rho[{\bf r}''-{\bf r}_0(t)] }{|{\bf r}-{\bf r}''| }\dd {\bf r}''=\frac{q}{\left[ {\bf r}-{\bf r}_0(t)\right]^2}\neq 0\,.
\end{split}
\end{equation}
Since determination of the function $\Psi$ for the velocity gauge as $\grad\Psi=\left(-\grad\chi_{\rm C}\right)={\bf A}_{\rm C}-{\bf A}_{\rm L}$  is given for the Coulomb gauge is absent in Sec.~VII of~\cite{Jack} it cannot be accepted that the gauge function for transformations: 
$\left( \varphi^{(v)},\,{\bf A}^{(v)}\,\to\,\varphi_{\rm L},\,{\bf A}_{\rm L}\right)$ is given by Eq.~(7.5) of the cited paper.

Therefore, one can conclude that if such gauge functions exist, they should be found by means of more accurate mathematical procedure.

\section{Conclusions}

In this work, a problem of validity of different gauge fixings for application to electrodynamical systems is analyzed. Among the many different gauges, only three are applicable to the consideration of electrodynamic systems without any restrictions. But in two of these three gauges it is assumed that the scalar potential propagates with superluminal speed. It is shown in this work that this property is a source of pathological solutions for the electric field: calculation of the electric field in these gauges yields a solution predicting `superluminal' field propagation. These solutions are pathological because superluminal propagation of electromagnetic fields has not been detected.

In fact, the existence of a solution describing superluminal propagation can be demonstrated with a minimum of calculations. Since both scalar and vector potentials exhibit superluminal propagation (the latter exhibits this type of propagation due to its source), a superluminal electric field will be absent in the system if the derivatives of the two potentials cancel each other out. However, the vector potential is proportional to the velocity of the charge creating the potentials, while the scalar potential has no such dependence. Thus, mutual cancellation is impossible. Moreover, the absence of this cancellation is confirmed by direct calculations of the electric field in the `superluminal sector' (calculations for the Coulomb gauge in Sec. 2).

Finally it can be concluded that, to avoid appearance of possible pathological solutions, the only gauge, {\it i.e.} of Lorenz, should be used in the applied electrodynamical calculations.

\end{document}